\documentclass[superscriptaddress,12pt]{revtex4-1}

\usepackage{xcolor}
\usepackage{physics}
\usepackage{graphicx}
\usepackage{dcolumn}
\usepackage{bm}
\usepackage{footmisc}

\usepackage{mathtools}
\usepackage{float}

\begin{document}

\title{Higher-dimensional spin selectivity in chiral crystals}

\author{Yinong Zhou}
\affiliation{%
 Department of Physics and Astronomy, University of California, Irvine, California
92697-4575, USA\\
 }%

\author{Dmitri Leo M. Cordova}
\affiliation{%
 Department of Chemistry, University of California, Irvine, California
92697-4575, USA\\
 }%

\author{Griffin M. Milligan}
\affiliation{%
 Department of Chemistry, University of California, Irvine, California
92697-4575, USA\\
 }%

\author{Maxx Q. Arguilla}
\affiliation{%
 Department of Chemistry, University of California, Irvine, California
92697-4575, USA\\
 }%

\author{Ruqian Wu}%
\email{wur@uci.edu}
\affiliation{%
	Department of Physics and Astronomy, University of California, Irvine, California
92697-4575, USA\\
}%

\begin{abstract}
\textbf{Abstract:} This study aims to investigate the interplay between chiral-induced spin-orbit coupling along the screw axis and antisymmetric spin-orbit coupling (ASOC) in the normal plane within a chiral crystal, using both general model analysis and first-principles simulations of InSeI, a chiral van der Waals crystal. While chiral molecules of light atoms typically exhibit spin selectivity only along the screw axis, chiral crystals with heavier atoms can have strong ASOC effects that influence spin-momentum locking in all directions. The resulting phase diagram of spin texture shows the potential for controlling phase transition and flipping spin by reducing symmetry through surface cleavage, thickness reduction or strain. We also experimentally synthesized high-quality InSeI crystals of the thermodynamically stable achiral analogue which showed exposed (110) facets corresponding to single-handed helices to demonstrate the potential of material realization for higher-dimensional spin selectivity in the development of spintronic devices.
\end{abstract}

\maketitle

Chiral structures, which refers to the geometric characteristic of a rigid object or system that prevents it from being superimposed onto its mirror image, are found abundantly in nature, from the swirling designs of galaxies to the molecular underpinnings of life \cite{lough2002chirality,wagniere2007chirality,cho2023bioinspired}. Chiral materials can exist on various scales, encompassing not only molecules but also larger structures, including specific chiral crystals \cite{flack2003chiral} and metamaterials \cite{wang2009chiral,wang2016optical,fernandez2019new}. The importance of chirality cannot be understated, as it can dictate a substance's optical, magnetic, and electronic properties, such as magneto-chiral dichroism \cite{rikken1997observation,berova2000circular}, magnetic skyrmion \cite{bogdanov1994thermodynamically,nagaosa2013topological}, and electrical magnetochiral anisotropy \cite{rikken2001electrical,krstic2002magneto,pop2014electrical}, paving the way for diverse applications. One such application is in the field of spintronics \cite{hirohata2020review}, where the phenomenon of chiral-induced spin selectivity (CISS) comes into play \cite{naaman2012chiral,naaman2015spintronics,dalum2019theory,naaman2019chiral,fransson2022chiral}. As electrons traverse a chiral system, those with a specific spin orientation are transmitted more efficiently than their counterparts with the opposite spin. The ability to harness electron spin in spintronics has the potential to expedite data transfer, reduce power usage, and augment memory and processing capabilities \cite{hirohata2014future,bandyopadhyay2015introduction,nanotechnology2015memory}.

The coupling of electronic motion and spin, known as the spin-orbit coupling (SOC) effect, plays a crucial role in creating a chiral spin filter through the CISS \cite{naaman2015spintronics,dalum2019theory,naaman2019chiral}. Studies on chiral molecules, such as the archetypal double helix DNA \cite{gohler2011spin,du2020vibration,naaman2022chiral}, have demonstrated that the inherent chiral potential induces a substantial SOC which, in turn, yields significant spin polarization along the screw axis \cite{gutierrez2012spin,guo2012spin,medina2012chiral}. For an electron moving in a single helix chain with a momentum $\bm{p}$, the SOC can be expressed as $H_{SOC}=\frac{\hbar}{4m^2c^2} \nabla V \cdot (\bm{\sigma} \cross \bm{p})$, where $\hbar$ is the reduced Planck's constant, $m$ is the electron mass, $c$ is the speed of light, $V$ is the electrostatic potential, and $\bm{\sigma}=(\sigma_{x},\sigma_{y},\sigma_{z})$ represents the Pauli matrices. In one-dimensional (1D) chains, the terms containing $\sigma_{x}$ and $\sigma_{y}$ are canceled by oscillation, leaving only the $\sigma_{z}$ term to contribute to the spin polarization along the screw axis. This results in a chiral-induced SOC (CISOC) $H_c=\lambda_C k_z \sigma_z$, where $\lambda_C$ denotes the strength of CISOC \cite{yu2020chirality}. 

To broaden the potential applications of chirality and deepen our understanding of its fundamental science, it is more intriguing to explore chiral crystals which extend periodicity in all three dimensions. This introduces additional complexities and opportunities to study new chirality-induced phenomena. Firstly, one could anticipate the development of spin current in all three dimensions, offering greater flexibility in device design. Furthermore, compared to chiral molecules, the SOC effect in chiral crystals can be considerably stronger and adjustable by manipulating symmetry and incorporating heavy atoms. Prior research on chiral crystals has largely overlooked the spin-filtering aspect, instead focusing on Kramers-Weyl fermions formed at time-reversal-invariant momenta in metallic states or minimal bandgaps \cite{xu2015discovery,chang2018topological,sanchez2019topological,li2019chiral}. Therefore, investigating and controlling spin selectivity in chiral crystals could pave the way for a new research direction. This would facilitate the development of diverse applications in technology and materials science, further expanding the impact of chirality in these fields.

In recent decades, there has been considerable interest in crystals with 1D sub-units that are held together by weak interactions such as vdW or ionic bonding \cite{stolyarov2016breakdown,liu2017low,zhang2018thermal,zhu2021spectrum}. Many of them exhibit structural chirality within each 1D unit, for example, tellurium nanowires \cite{kramer2020tellurium,calavalle2022gate}, tantalum disilicide \cite{shiota2021chirality}, and monoaxial chiral dichalcogenide \cite{inui2020chirality,nabei2020current}. Typically, the CISOC strength in these materials is substantially larger than the antisymmetric SOC (ASOC) strength, leading to the omission of the ASOC effect normal to the screw axis in previous studies \cite{roy2022long}. However, the ASOC becomes particularly important for some crystals, such as indium selenoiodide (InSeI), which consists of a bundle of atomically precise 1D helical chains with heavy elements and a huge twist angle (135$^\circ$) \cite{sawitzki1980kristallstrukturen,jiang2020computational,jiang2020computational,choi2022one}. The ASOC becomes prominent in these crystals, and the interplay with the CISOC effect may give rise to unique spin configurations in the three-dimensional (3D) Brillouin zone.

In this letter, we investigate the interplay of CISOC and ASOC and establish a phase diagram of the 3D spin texture with varying four ASOC contributions. When Weyl or Dresselhaus SOC dominates in a chiral crystal, it may lead to a high-spin-coherency phase which has spin selectivity in three directions. This interesting finding is verified by first-principles calculations on a chiral helical van der Waals (vdW) crystal, indium selenoiodide (InSeI), whose first valence band exhibits high spin coherency with the dominating Weyl ASOC. Furthermore, we find that the ASOC is sensitive to the change of structural symmetry, and a phase transition with spin-flipping can be induced by symmetry reduction. As a major step towards the realization, we experimentally show that a chiral (110) surface can be exposed in the thermodynamically stable achiral analogue of InSeI.

To give a general picture for the interplay of ASOC and CISOC, we write a $\textit{\textbf{k}}\cdot\textit{\textbf{p}}$ effective Hamiltonian of a chiral crystal with the screw axis along $z$ direction.
 \begin{equation}
    \begin{split}
 	H &= H_R + H_W + H_D + H_{D'} + H_C\\
    &=\lambda_R(\sigma_x k_y-\sigma_y k_x)+\lambda_W(\sigma_x k_x+\sigma_y k_y)\\
    &+\lambda_D(\sigma_x k_x-\sigma_y k_y)+\lambda_{D'}(\sigma_x k_y+\sigma_y k_x)+\lambda_C k_z \sigma_z,
    \end{split}
 \end{equation}
where $\lambda_i$ ($i=R, W, D, {D'}, C$) represents the strength of each SOC term. The first four terms are ASOCs including Rashba SOC ($H_R$) with tangential spin texture $\Vec S=(k_y,-\,k_x,0)$; Weyl SOC ($H_W$) with radial spin texture $\Vec S=(k_x,k_y,0)$; and two Dresselhaus SOCs ($H_D$ and $H_{D'}$) with tangential-radial spin textures $\Vec S=(k_x,-k_y,0)$ and $\Vec S=(k_y,k_x,0)$, respectively. The last term is the CISOC ($H_C$), where the spin texture can be expressed as $\Vec S=(0,0,k_z)$. Solving the Hamiltonian in Eq. (1), we can get analytical solutions of spin texture for the spin-up state as $\Vec S=(S^x,S^y,S^z)$, where:
\begin{equation}
    \begin{split}
    &S^x=\mp \frac{2}{N^2}\frac{(\lambda_W+\lambda_D)k_x+(\lambda_R+\lambda_{D'})k_y}{\lambda_C\textit{$k_z$} \pm \mathcal{E}},\\
    &S^y=\mp \frac{2}{N^2}\frac{(\lambda_{D'}-\lambda_R)k_x+(\lambda_W-\lambda_D)k_y}{\lambda_C\textit{$k_z$} \pm \mathcal{E}},\\
    &S^z=\mp \frac{2}{N^2}\frac{\lambda_C\textit{$k_z$}}{\lambda_C\textit{$k_z$} \pm \mathcal{E}},
    \end{split}
\end{equation}
for $k_z\ge0$ and $k_z<0$, respectively. $\mathcal{E}$ represents the absolute value of the eigenvalues of $H$. More detailed derivations are shown in Section I of the Supplemental Material \footnote{See Supplemental Material at http://link.aps.org/supplemental/xxx, for more details which include Ref.\cite{blochl1994projector,kresse1999ultrasoft,perdew1996generalized,kresse1996efficient,methfessel1989high,grimme2010consistent,inui2020chirality,valenzuela2006direct,saitoh2006conversion,kimura2007room,wang2014scaling,niimi2015reciprocal}}.

\begin{figure*}
	\centering
	\includegraphics[width=1\textwidth]{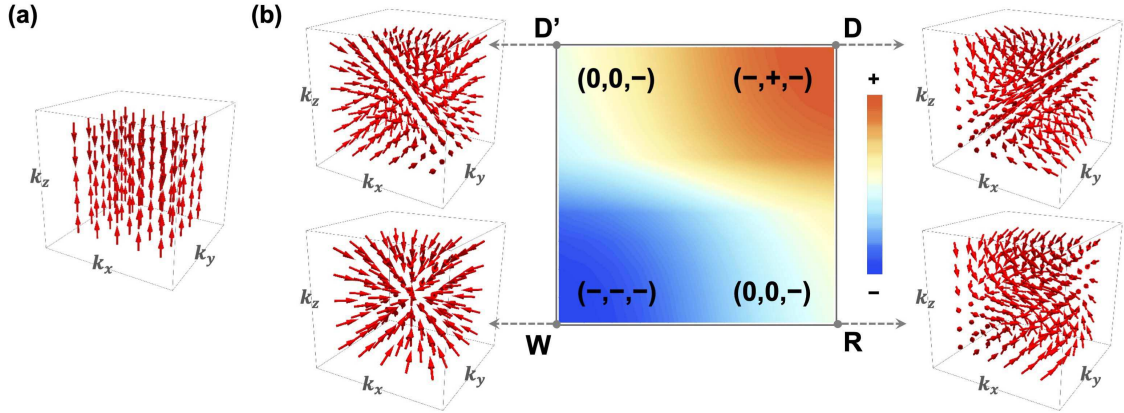}
	\caption{\label{fig:epsart} (a) 3D spin texture of the CISOC. (b) Phase diagram of the interplay of CISOC with different ASOC contributions including the Rashba (R), Weyl (W), and two types of Dresselhaus (D and D') SOCs. The phases of $(-\,,-\,,-\,)$, $(-\,,+,-\,)$, and $(0,0,-\,)$ represent the sign of spin coherency in three directions,  where $+$ and $-\,$ represent the spin is parallel or antiparallel with the current direction, and 0 represents $+$ and $-\,$ spin are canceled out. The color represents the strength of the spin coherency. The typical 3D spin textures are plotted at four corners corresponding to the grey dots in the phase diagram.}
\end{figure*}

Accordingly, the 3D spin texture is plotted in Fig. 1. With only CISOC, the spin orientation always points to $-\hat{k}_z$ direction, as shown in Fig. 1(a), leading to the spin selectivity in $k_z$ direction. However, additional ASOC terms lead to much more complex spin textures, and spin selectivity can be realized in all three dimensions. To visualize this behavior, we construct a phase diagram of the 3D spin selectivity in Fig. 1(b) with the different contributions of ASOCs cooperating with a consistent CISOC strength. The phases are defined by extracting the sign of the spin coherency $(h_x,h_y,h_z)$ for the spin current in positive $k$ directions, where 
\begin{equation}
    h_x=\sum_{\substack{k_x>0,\\k_y,k_z}} S^x(\vec k),\,\,\,
    h_y=\sum_{\substack{k_y>0,\\k_x,k_z}} S^y(\vec k),\,\,\,
    h_z=\sum_{\substack{k_z>0,\\k_x,k_y}} S^z(\vec k).\,\,\,
\end{equation}
For instance, with only Weyl and CISOC $\lambda_W=\lambda_C$ (other $\lambda_i=0$), the spin coherency is negative in all three directions, leading to a $(-\,,-\,,-\,)$ phase. For $\lambda_D=\lambda_C$ (other $\lambda_i=0$), the spin coherency is positive in $k_y$ direction but negative in $k_x$ and $k_z$ directions, resulting a $(-\,,+,-\,)$ phase. For $\lambda_R=\lambda_C$ or $\lambda_{D'}=\lambda_C$ (other $\lambda_i=0$), the spin coherency is canceled out due to the tangential spin texture in $k_x$ and $k_y$ directions; but in $k_z$ direction, the spin selectivity survives due to the structure chirality, resulting in a $(0,0,-\,)$ phase. More detailed discussions of the phase diagram can be found in Section II of the Supplemental Material \footnote[1]{}.

\begin{figure}
	\centering
	\includegraphics[width=0.6\columnwidth]{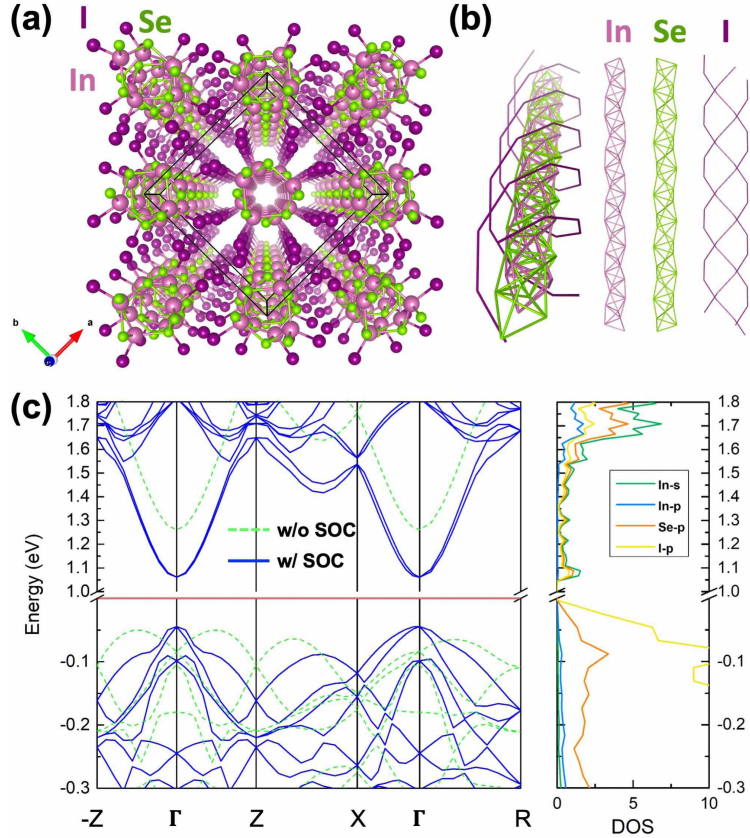}
	\caption{\label{fig:epsart} Lattice and band structure of InSeI. (a) The bulk crystal with right-handed chirality. (b) The 1D nanochain with the decomposed chains for In, Se, and I, respectively. (c) The band structures of the bulk in (a) with and without SOC and the orbital projected DOS with the green, blue, orange, and yellow lines represent In-\textit{s}, In-\textit{p}, Se-\textit{p}, and I-\textit{p} orbitals, respectively.}
\end{figure}

As the ASOCs can be tuned by reducing symmetry through surface cleavage, thickness reduction or strain, the possibility of having a rich spin texture and spin selectivity in a single system is very intriguing for the development of spintronic devices. For the realization of these results, one needs to find chiral crystals with large ASOCs. As shown in Fig. 2(a,b), we construct a chiral InSeI crystal formed by 1D helical chains with the [InISe$_3$]$_n$ tetrahedral motif wherein each distinct elemental site following a tetrahelix defined by a large 135$^\circ$ twist angle. These helical chains are stacked through weak vdW interactions. As shown in Fig. 2(a), the bulk InSeI with space group $P4_3$ shows the right-handed chirality with a screw symmetry including a $C_4$ rotational symmetry in $xy$ plane and a $c/4$ translational symmetry along $z$ direction, where $c$ represents the lattice constant along $z$ direction. Aside from the structural chirality, all elements in InSeI show a strong intrinsic SOC strength, especially in the heavier I atoms \cite{wittel1974atomic}. With the SOC included, the electronic band structure from the density functional calculations (DFT) shows spin splitting both in the directions along and normal to the screw axis, as shown in Fig. 2(c). The valence band splitting is larger than that of conduction bands because the valence bands are mainly contributed by I-\textit{p} orbitals with large SOC strength, as also reflected by the projected density of states (DOS) shown in Fig. 2(c). The calculation method and parameters are described in Section III of the Supplemental Material \footnote[1]{}.

\begin{figure}
	\centering
	\includegraphics[width=0.6\columnwidth]{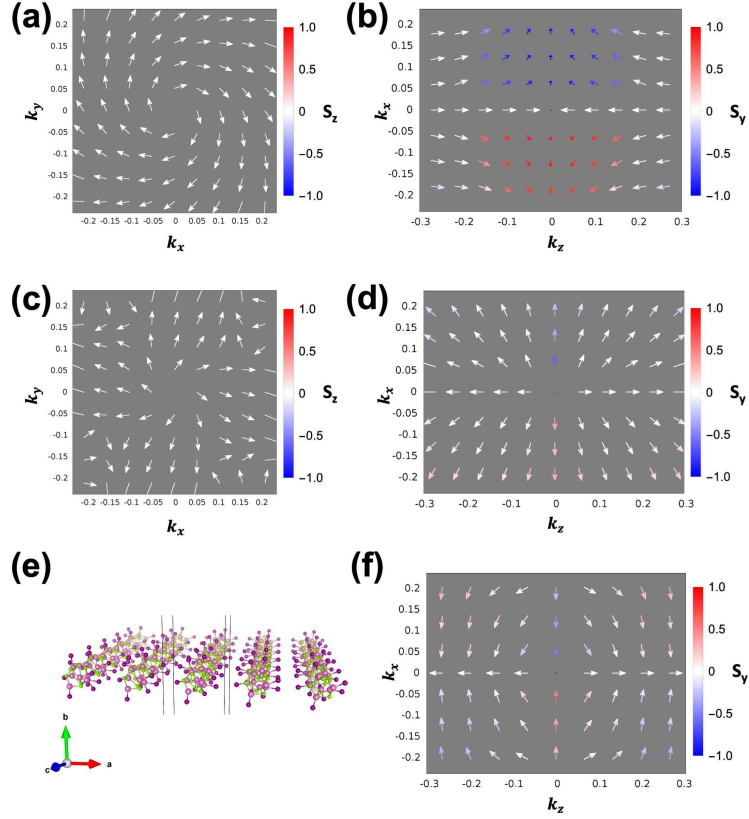}
	\caption{\label{fig:epsart} The spin texture of the first (a,b) conduction and (c,d) valence band in (a,c) $k_x$-$k_y$ plane and (b,d) $k_z$-$k_x$ plane, respectively. The color of the spin vectors represents the out-of-plane spin component. (e) The lattice structure of the monolayer for the (100) surface of the bulk in Fig. 2(a). (f) The spin texture of the first valence band of the monolayer in (e).}
\end{figure}

To investigate spin configurations, we plot the spin texture for the first conduction and valence band, as shown in Fig. 3 (a-d). As the bulk InSeI has the $C_4$ rotational symmetry in the normal plane, i.e., the $k_x$-$k_y$ plane, only Rashba and Weyl ASOC are allowed \cite{zhang2022topological}. For the first conduction band, the spin in the $k_x$-$k_y$ plane shows a tangential texture, corresponding to a Rashba-dominant ASOC [see Fig. 3(a)]. In contrast, the first valence band shows a radial spin texture, corresponding to a Weyl-dominant ASOC [see Fig. 3(c)]. Through the interplay with the CISOC in $k_z$ direction, the conduction band shows a tangential-radial spin texture in the $k_z$-$k_x$ plane [see Fig. 3(b)], while the valence band shows a radial spin texture [see Fig. 3(d)]. Importantly, we may extract SOC strength by fitting the spin textures obtained from the model and first-principle calculations, as $\lambda_R/\lambda_W=-\,S^y/S^x$ for $k_y=k_z=0$; and $\lambda_W/\lambda_C=S^xk_z/(S^zk_x)$ for $k_y=0$. Using $\lambda_C=C$ ($C$ is a constant) as the unit, we get $\lambda_W=-\,0.5\lambda_C$ and $\lambda_R=-\,1.2\lambda_C$ for the first conduction band [Fig. 3(a,b)]. For the first valence band [Fig. 3(c,d)], we get $\lambda_W=1.5\lambda_C$ and $\lambda_R=0.7\lambda_C$. Details of the fitting procedure and the opposite spin textures for the second conduction and valence bands are discussed in Section IV of the Supplemental Material \footnote[1]{}. Considering the spin coherency in 3D, the conduction band corresponds to the Rashba-dominant $(0,0,-\,)$ phase in Fig. 1(b) with only one direction spin selectivity in $k_z$ direction. The valence band corresponds to the Weyl-dominant $(+,+,+)$ phase with the spin selectivity in all three directions, where the $+$ sign indicates that the first valence band corresponds to a spin-down state with higher energy. The phase diagram of the spin-down state with the opposite sign of the Dresselhaus SOC is discussed in Fig. S3 and S4 in Section II of the Supplemental Material \footnote[1]{}.

Because ASOC is sensitive to crystalline symmetry, we explore the possibility of introducing the Dresselhaus SOC in InSeI by reducing its symmetry to $C_2$. Different from $C_4$ symmetry that only has Rashba and Weyl ASOCs, the $C_2$ symmetry allows all four ASOCs \cite{zhang2022topological}. We constructed a monolayer of nano-chains in the (100) plane, as shown in Fig 3(e). The spin texture for the first valence band becomes a tangential-radial type in the $k_z$-$k_x$ plane [see Fig. 3(f)]. By fitting the spin texture around $\Gamma$ point, we find that except Rashba and Weyl SOC, Dresselhaus SOC terms are introduced with $\lambda_D=-\,2.7\lambda_C$ and $\lambda_{D'}=0.4\lambda_C$ \footnote[1]{}. The strong Dresselhaus SOC leads to a phase transition from the Weyl-dominant $(+,+,+)$ phase to the Dresselhaus-dominant $(-\,,+,+)$ phase with the opposite spin selectivity in $k_x$ direction, comparing Fig. 3(d) and (f). The phase transition is labeled in the spin-down phase diagram in Fig. S3 \footnote[1]{}. The phase transition can also be induced by applying compressive strain on the bulk to reduce the symmetry. A gradual spin-flipping of the spin texture along $k_x$ direction can be found in Fig. S9 in Section IV of the Supplemental Material \footnote[1]{}.

\begin{figure}
	\centering
	\includegraphics[width=0.6\columnwidth]{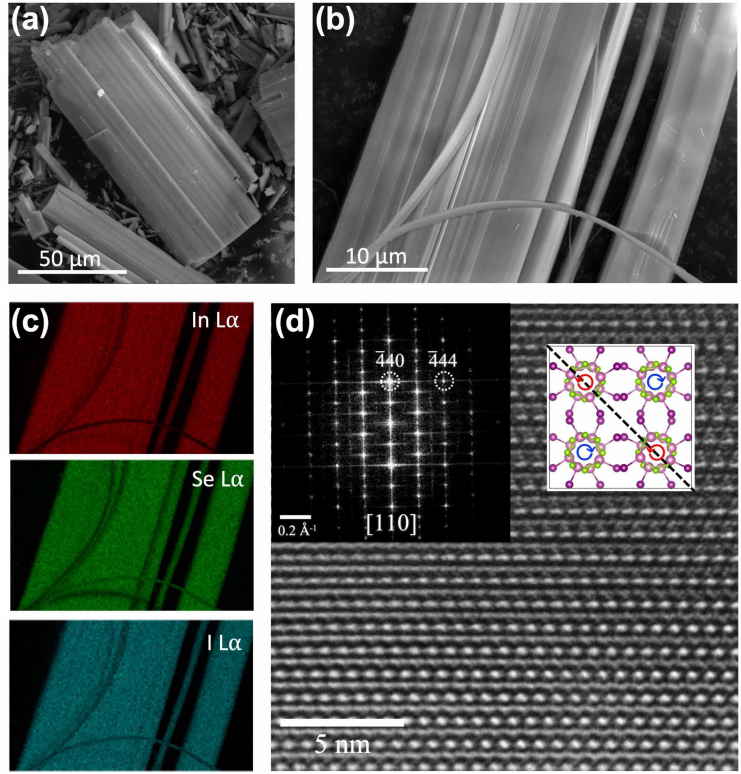}
	\caption{\label{fig:epsart} (a,b) SEM images of InSeI. (c) EDS elemental mapping of L$\alpha$ representing the relaxation from M ($n=3$) to L ($n=2$) electrons of In, Se, and I, respectively. (d) TEM image for the (110) surface with the FFT pattern inserted on the top left. The top view of the crystal unit cell is inserted on the top-right, where the black dashed line represents the (110) surface with the same chirality. The red and blue circles with arrows in the insert structure represent left- and right-handed nanochains, respectively.} 
\end{figure}

Most excitingly, we were able to synthesize sizeable high-quality crystals of the reported thermodynamically stable analogue of InSeI \cite{sawitzki1980kristallstrukturen,choi2022one}. The synthesis method can be found in Section V of the Supplemental Material \footnote[1]{}. As shown in the scanning electron microscopy (SEM) images in Fig. 4(a,b), the air- and temperature-stable sample displays a cleavable stepped surface which reflects the 1D vdW structure in the bulk. The energy-dispersive spectroscopy (EDS) shows the purity and uniform distribution of the constituent In, Se, and I elements in the structure [Fig. 4(c)]. However, it is important to note that the experimentally synthesized InSeI crystallizes in the achiral space group $I4_1/a$ [42, 44], which contains both handedness (left- and right-handedness) of the helices, as shown in the insert on the top-right of Fig. 4(d). Encouragingly, owing to the vdW interaction between the helical chains and the intrinsic size of the crystals, InSeI crystals are readily cleavable and orientable in select facets parallel to the long chain direction, as shown in Fig. 4(b). These properties of the bulk crystals allow for the exposure of a (110) surface which corresponds to a surface comprised of helices with only one type of handedness, as the dashed line shown in the inserts of Fig. 4(d). High-resolution transmission electron microscopy (HRTEM) imaging and the corresponding indexing of the fast Fourier transform (FFT, inset) of an exposed cleaved facet of InSeI show that it is possible to selectively cleave and expose the sought-after (110) surface of the achiral InSeI crystal, as is shown in Fig. 4(d). The corresponding calculation of the spin texture for the (110) surface can be found in Section IV of the Supplemental Material [37]. As expected, these calculations confirm that the (110) facet which bears the helices with single handedness, in the monolayer regime, also involves strong Dresselhaus terms as discussed in Fig. 3(e,f). The design of a spintronic device to measure spin selectivity in three directions is discussed in Section VI of the Supplemental Material \footnote[1]{}.
 
In conclusion, we demonstrated that the 3D spin texture can be attained with the combination of out-of-plane CISOC and in-plane ASOC. With the interplay of different types of ASOC, a general phase diagram with different spin coherency phases is generated by the model studies. When Weyl or Dresselhaus SOC dominates, the spin texture exhibits a 3D spin selectivity arising from the spin-momentum locking. The modeling results were verified by DFT calculations for InSeI, a chiral vdW crystal. A large spin splitting and radial spin texture were found for the first valence band, as a first example of the realization of the 3D spin selectivity in actual materials. By reducing the symmetry, a strong Dresselhaus SOC is produced in the monolayer InSeI, leading to a phase transition with an opposite spin selectivity along the $k_x$ direction. Remarkably, we were able to experimentally synthesize the thermodynamically stable achiral InSeI analogue in high quality and demonstrated that the (110) surface shows the aligned helices with the same handedness. More experiments of the spin measurement using magnetic conductive probe atomic force microscopy \cite{kiran2016helicenes} and the device applications are in progress. 

Our findings indicate that the ASOC strength can play an important role in the chiral crystal with heavier elements, like InSeI we discussed herein. The interplay of the CISOC along the screw axis and ASOC in the normal plane leads to different phases of spin texture with high spin coherency in all three directions. The higher dimensional spin selectivity offers numerous advantages including greater control over the spin states, the potential for new functionalities that are not possible with monodirectional selectivity, improved performance in certain applications, and increased flexibility in designing and implementing spin-based technologies. We anticipate that these advancements will significantly broaden the development of quantum computing, spintronics, spin filtering, and other applications that rely on the manipulation of spin states.

Y. Zhou and R. Wu were supported by DOE-BES (Grant No. DE-FG02-05ER46237). Computational simulations were performed at the U.S. Department of Energy Supercomputer Facility (NERSC). D. L. M. Cordova, G. M. Milligan, and M. Q. Arguilla acknowledge the UC Irvine Materials Research Institute (IMRI) for instrumental support. Facilities and instrumentation at IMRI are supported, in part, by the National Science Foundation through the UC Irvine Materials Research Science and Engineering Center grant number DMR-2011967.

\pagebreak

\hfill \break
{\large Supplemental material for}
\\
\begin{center}
	\textbf{\large Higher-dimensional spin selectivity in chiral crystals}
\end{center}

\hfill \break
\begin{center}
	{Yinong Zhou$^{1}$, Dmitri Leo M. Cordova$^{2}$, Griffin M. Milligan$^{2}$, Maxx Q. Arguilla$^{2}$ and Ruqian Wu$^{1,*}$}
\end{center}

\begin{center}
	{$^{1}$\textit{Department of Physics and Astronomy, University of California, Irvine, California 92697-4575, USA}}
\end{center}

\begin{center}
	{$^{2}$\textit{Department of Chemistry, University of California, Irvine, California
92697-4575, USA}}
\end{center}

\begin{center}
	{$^{*}$wur@uci.edu}
\end{center}

\clearpage

\def\thesection{\Roman{section}}

\setcounter{equation}{0}
\setcounter{figure}{0}
\setcounter{table}{0}
\setcounter{page}{1}

\makeatletter
\renewcommand{\theequation}{S\arabic{equation}}
\renewcommand{\thefigure}{S\arabic{figure}}

\hfill \break

\tableofcontents

\clearpage

\section{Derivation of SOC Hamiltonian}
In this section, we show the detailed derivation of the spin-orbit coupling (SOC) Hamiltonian in Eq. (1) in the main text. We rewrite the Hamiltonian below for convenience:
\begin{equation}
    \begin{split}
 	&H = H_R + H_W + H_D + H_{D'} + H_C;\\
 	&H_R = \lambda_R[\sigma_x k_y-\sigma_y k_x],\\
    &H_W = \lambda_W[\sigma_x k_x+\sigma_y k_y],\\ 
    &H_D = \lambda_D[\sigma_x k_x-\sigma_y k_y],\\ 
    &H_{D'} = \lambda_{D'}[\sigma_x k_y+\sigma_y k_x],\\ 
    &H_C = \lambda_C k_z \sigma_z.
    \end{split}
\end{equation}
The first four terms are antisymmetric SOC (ASOC), whose spin texture is shown in Fig. S1. The SOC strength is expressed as $\lambda_R$ for Rashba, $\lambda_W$ for Weyl, $\lambda_D$ and $\lambda_{D'}$ for Dresselhaus SOC. The last term represents chiral-induced SOC (CISOC) with the SOC strength $\lambda_C$. The sign of the $\lambda_C$ is determined by the material’s helicity. In the following derivation, we consider $\lambda_C$ to be positive for the right-handed chiral crystal.

\begin{figure}[b]
	\centering
	\includegraphics[width=1\columnwidth]{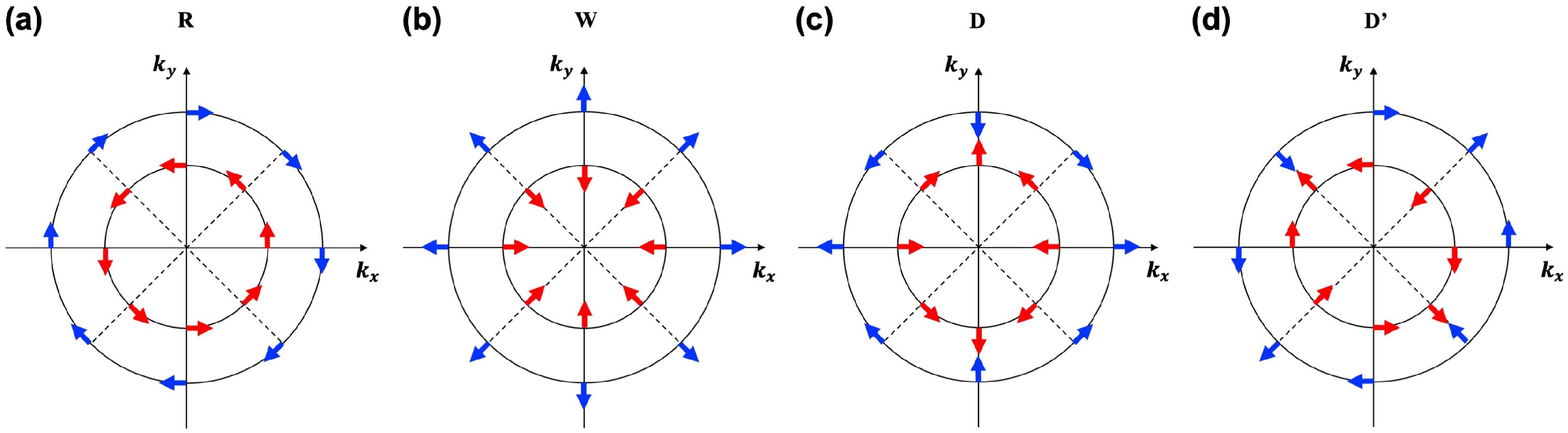}
	\caption{\label{fig:epsart} Spin texture of ASOC: (a) Rashba (R) with $\Vec S=(k_y,-\,k_x,0)$; (b) Weyl (W) with $\Vec S=(k_x,k_y,0)$ (c,d) Dresselhaus (D and D') with $\Vec S=(k_x,-k_y,0)$ and $\Vec S=(k_y,k_x,0)$, respectively.}
\end{figure}

The Hamiltonian in Eq. (1) can be expanded as
\begin{equation}\tag{S1}
    \begin{split}
    H = 
    \begin{pmatrix}
    \textit{$\lambda_C$} k_z & \textit{$h_{12}$}\\
    \textit{$h_{12}$}^* &-\textit{$\lambda_C$} k_z
    \end{pmatrix},
    \end{split}
\end{equation}
where
\begin{equation}\tag{S2}
    \textit{$h_{12}$} = [(\lambda_W+\lambda_D)+\textit{$i$}(\lambda_R-\lambda_{D'})] k_x
    + [(\lambda_R+\lambda_{D'})-\textit{$i$}(\lambda_W-\lambda_D)] k_y.
\end{equation}
Solving Eq. (S1), we can obtain the eigenvalues:
\begin{equation}\tag{S3}
    E_{1,2} = \mp \sqrt{{\lambda_C}^2 {k_z}^2 + \textit{$h_{12}$} \textit{$h_{12}$}^*} \equiv \mp \mathcal{E},
\end{equation}
where $1,2$ represents the spin-up and spin-down states, respectively. $\mathcal{E}$ represents the absolute value of the eigenvalues. The corresponding eigenvectors are
\begin{equation}\tag{S4}
    \begin{split}
    v_1=\begin{dcases}
    \{-\frac{\textit{$h_{12}$}}{{\lambda_C}\textit{$k_z$} + \mathcal{E}},1\}, & \text{if $k_z\ge0$};\\
    \{1 , \frac{\textit{$h_{12}$}^*}{{\lambda_C}\textit{$k_z$} - \mathcal{E}}\}, & \text{if $k_z<0$};
    \end{dcases}\\
    v_2=\begin{dcases}
    \{1 , \frac{\textit{$h_{12}$}^*}{{\lambda_C}\textit{$k_z$} + \mathcal{E}}\}, & \text{if $k_z\ge0$};\\
    \{-\frac{\textit{$h_{12}$}}{{\lambda_C}\textit{$k_z$} - \mathcal{E}},1\}, & \text{if $k_z<0$}.
    \end{dcases}
    \end{split}
\end{equation}
The spin texture $\Vec S=(S^x,S^y,S^z)$ of the spin-polarized states are
\begin{equation}\tag{S5}
    \begin{split}
    S_{1,2}^x &= \bra{\frac{v_{1,2}}{N}} \sigma_x \ket{\frac{v_{1,2}}{N}}\\
    &=\begin{dcases}\mp \frac{2}{N^2}\frac{(\lambda_W+\lambda_D)k_x+(\lambda_R+\lambda_{D'})k_y}{{\lambda_C}\textit{$k_z$} + \mathcal{E}}, & \text{if $k_z\ge0$};\\
    \pm \frac{2}{N^2}\frac{(\lambda_W+\lambda_D)k_x+(\lambda_R+\lambda_{D'})k_y}{{\lambda_C}\textit{$k_z$} - \mathcal{E}}, & \text{if $k_z<0$};\\
    \end{dcases}\\
    S_{1,2}^y &= \bra{\frac{v_{1,2}}{N}} \sigma_y \ket{\frac{v_{1,2}}{N}}\\
    &=\begin{dcases}\mp \frac{2}{N^2}\frac{(\lambda_{D'}-\lambda_R)k_x+(\lambda_W-\lambda_D)k_y}{{\lambda_C}\textit{$k_z$} + \mathcal{E}}, & \text{if $k_z\ge0$};\\
    \pm \frac{2}{N^2}\frac{(\lambda_{D'}-\lambda_R)k_x+(\lambda_W-\lambda_D)k_y}{{\lambda_C}\textit{$k_z$} - \mathcal{E}}, & \text{if $k_z<0$};\\
    \end{dcases}\\
    S_{1,2}^z &=\bra{\frac{v_{1,2}}{N}} \sigma_z \ket{\frac{v_{1,2}}{N}}\\
    &=\begin{dcases}\mp \frac{2}{N^2}\frac{{\lambda_C}\textit{$k_z$}}{{\lambda_C}\textit{$k_z$} + \mathcal{E}}, & \text{if $k_z\ge0$};\\
    \pm \frac{2}{N^2}\frac{{\lambda_C}\textit{$k_z$}}{{\lambda_C}\textit{$k_z$} - \mathcal{E}}, & \text{if $k_z<0$}.\\
    \end{dcases}
    \end{split}
\end{equation}
Here, we use the piecewise functions to express the eigenvectors and spin textures to avoid dividing by zero when $\textit{$h_{12}$}=0$, i.e., $k_x=k_y=0$ or in the case of $\lambda_R=-\,\lambda_D$ and $\lambda_W=-\,\lambda_{D'}$.

\section{Phase diagram of spin coherency}

In the phase diagram in the main text of Fig. 1(b), the SOC strength of a point is defined as $\lambda_W=\lambda_C$, $\lambda_R=(a/b)\lambda_W$, $\lambda_{D'}=(c/d)\lambda_W$, $\lambda_D=(ac/bd)\lambda_W$, $a,b,c,d$ are the distance to the edges labeled in Fig. S2. The color scheme of Fig. 1(b) corresponds to the spin coherency in $k_y$ direction, i.e., $h_y$, to identify the difference between the phases. The phase diagram corresponding to $h_x$ is shown in Fig. S2, in which both Weyl- and Dresselhaus-dominant phases show a negative sign in $k_x$ direction. However, when either Rashba or Dresselhaus-D' SOC dominates, the spin coherency are canceled out due to the tangential spin texture in $k_x$ and $k_y$ directions, resulting in a $h_{x,y}\rightarrow0$.

In order to show the phase transition of the first valence band of InSeI, we plot a phase diagram of the spin-down state, i.e., $E_2$ in Eq. (S3), in Fig. S3. Obviously, the spin coherency for the spin-down state shows an opposite sign compared to the spin-up states, e.g., Weyl-dominant phase $(-\,,-\,,-\,)$ for spin-up state changes to phase $(+,+,+)$ for spin-down state. In addition, we find that the Dresselhaus SOC strength has the opposite sign to the other SOCs for InSeI. So, in the phase diagram of Fig. S3, we choose a negative sign for the $\lambda_D$. As a result, the Dresselhaus spin texture changes to $S=(-k_x,k_y,0)$, giving rise to the opposite spin along $k_x$ direction instead of along $k_y$ direction in Fig. 1(b), i.e., the Dresselhaus-dominant phase $(-\,,+,-\,)$ for spin-up state changes to phase $(-\,,+,+)$ for spin-down state with opposite $\lambda_D$. The bulk chiral InSeI [structure in Fig. 2(a)] corresponds to the star point in the Weyl-dominant $(+,+,+)$ phase, as labeled in Fig. S3. The monolayer for the (100) surface of the bulk [structure in Fig. 3(e)] corresponds to the triangle point in the Dresselhaus-dominant $(-\,,+,+)$ phase. The symmetry reducing from $C_4$ to $C_2$ leads a phase transition with a spin flipping in $k_x$ direction, comparing Fig. 3(d) and (f) in the main text.

We also plot the phase diagram of spin coherency of $h_y$ for the spin-down state in Fig. S4, which shows the opposite sign of spin coherency compared to spin-up state in Fig. S2.
\clearpage

\begin{figure}[H]
	\centering
	\includegraphics[width=0.4\columnwidth]{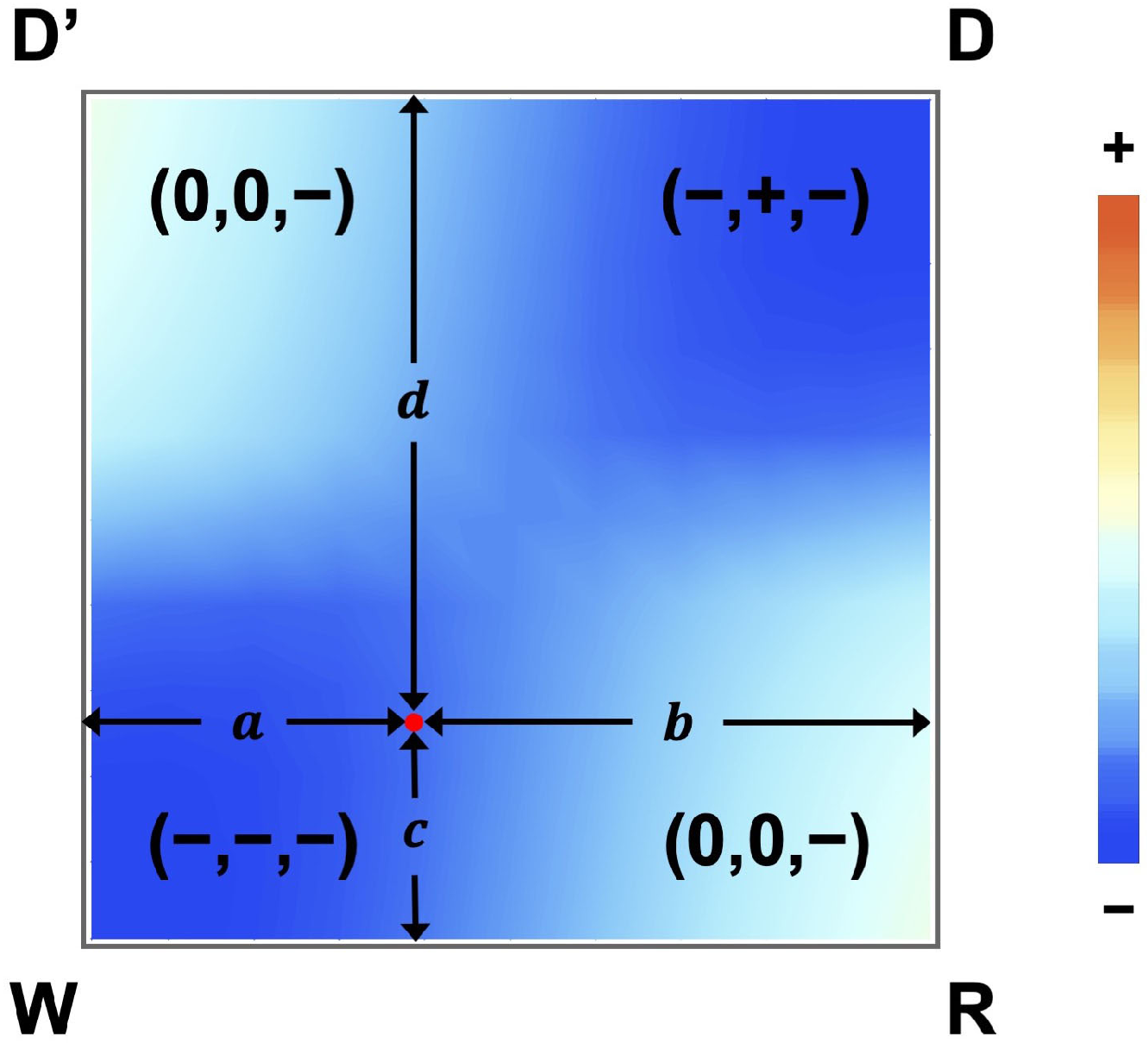}
	\caption{\label{fig:epsart} Phase diagram corresponding to spin coherency $h_x$ for the spin-up state, which shows the interplay of CISOC with different ASOC contributions including the Rashba (R), Weyl (W), and two types of Dresselhaus (D and D') SOCs. $a,b,c,d$ represent the distance from the red dot to the four edges used to define the SOC strength.}
\end{figure}

\begin{figure}[H]
	\centering
	\includegraphics[width=0.4\columnwidth]{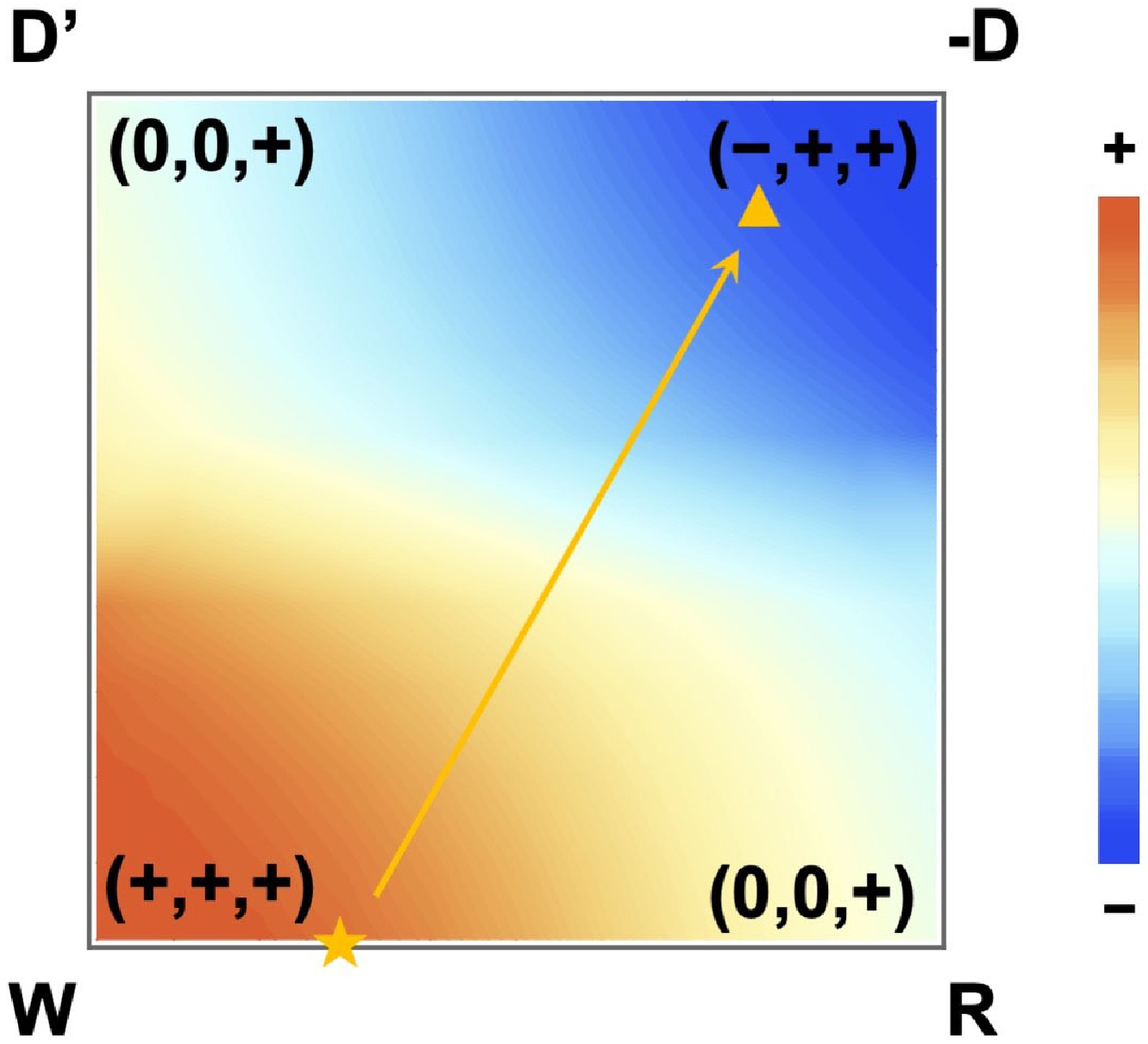}
	\caption{\label{fig:epsart} Phase diagram of spin coherency of $h_x$ for the spin-down state with the opposite sign of $\lambda_D$. The yellow star and triangle correspond to the first valence band of InSeI bulk in Fig. 2(a) and the monolayer in Fig. 3(e), respectively.}
\end{figure}

\begin{figure}[H]
	\centering
	\includegraphics[width=0.4\columnwidth]{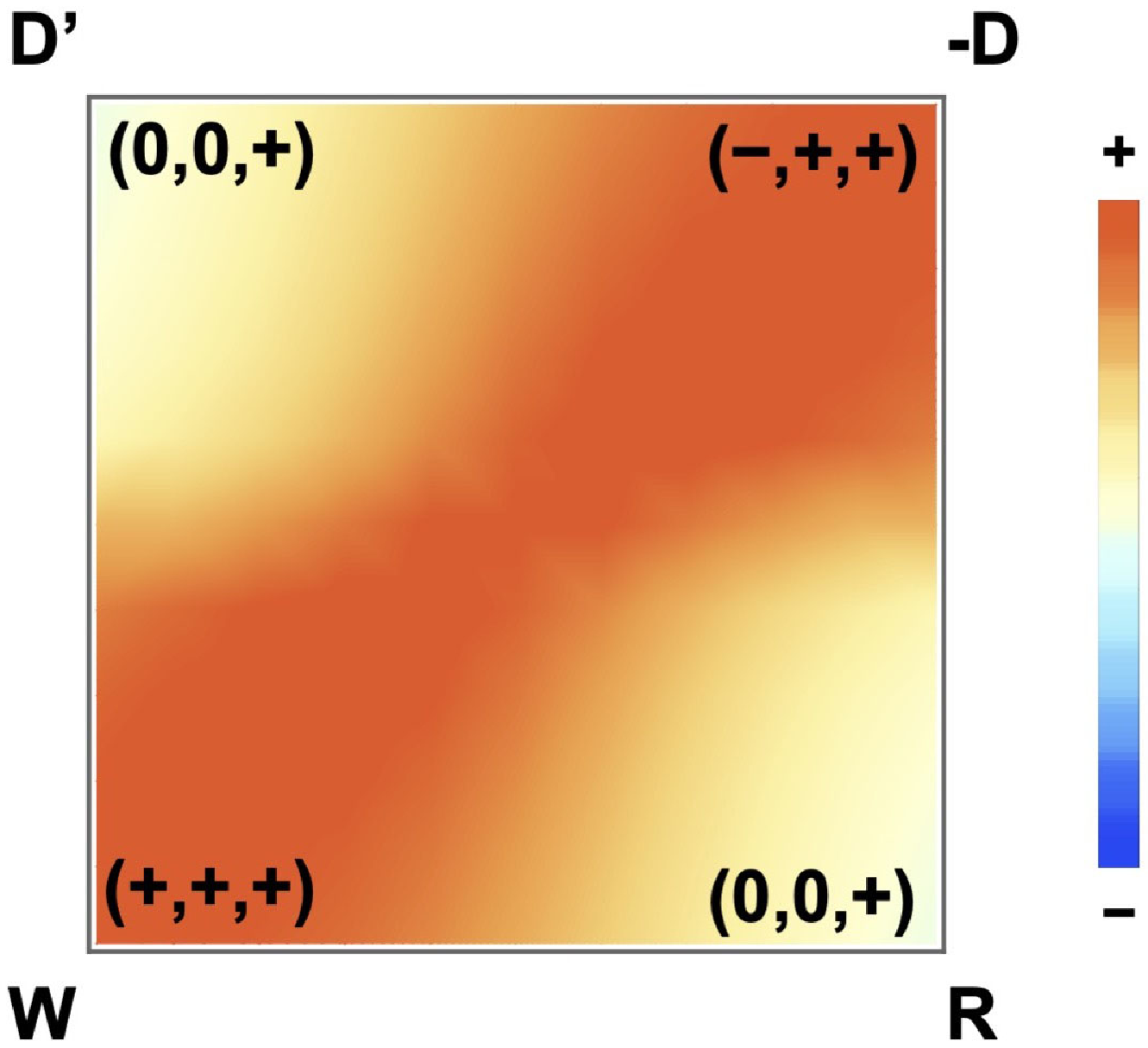}
	\caption{\label{fig:epsart} Phase diagram of spin coherency of $h_y$ for the spin-down state with the opposite sign of $\lambda_D$.}
\end{figure}

\newpage

\section{First-principles calculation method}

Our first-principles calculations are performed with the projector-augmented wave pseudopotentials \cite{blochl1994projector,kresse1999ultrasoft} and the generalized gradient approximation of Perdew-Burke-Ernzerhof \cite{perdew1996generalized} using Vienna Ab initio Simulation Package (VASP) \cite{kresse1996efficient} code. An energy cutoff of 450 eV and a 5$\cross$5$\cross$5 Monkhorst-Pack $k$-point grid is used \cite{methfessel1989high}. The structure is optimized until the atomic forces are smaller than 0.01 eV/\AA. For the vdW corrections, we used the DFT-D3 method \cite{grimme2010consistent}. The relaxed lattice constant for the chiral InSeI in Fig. 2(a) is $a=b=13.262$\AA, $c=10.398$\AA. For monolayer system, the lattice constant for the (110) surface of the chiral analogue is $a=10.398$\AA, $b=26.524$\AA. The vacuum layer is larger than 20\AA $\space$ to ensure decoupling between neighboring nanostructures. The spin texture is obtained by calculating expectation values of the Pauli matrix, $\expval{S^{\alpha}(\vec{k})}=\expval{\sigma_{\alpha}}{\Psi}$, $\alpha=x,y,z$. The spinor wavefunction at every point of the k-mesh grid is calculated by the VASP simulation package.

\section{More results about spin texture}

The spin texture for the first and second valence and conduction bands of chiral bulk InSeI are shown in Fig. S5 (a-d) in the $k_x$-$k_y$ plane and Fig. 6 (a-d) in the $k_z$-$k_x$ plane. The first and second valence or conduction band shows the opposite spin texture due to the energy splitting, comparing Fig. S5,6 (a) and (b) or (c) and (d). Then, we use the analytical solutions of the spin texture from the model study to fit the spin texture obtained by first-principle calculations. For the bulk InSeI with $C_4$ rotational symmetry in the $k_x$-$k_y$ plane, only Rashba and Weyl SOC are allowed, i.e., $\lambda_D=\lambda_{D'}=0$. Based on the analytical solution of the spin texture in Eq. (S5), we know that $S^y/S^x=-\,\lambda_R/\lambda_W$ for $k_y=k_z=0$; $S^xk_z/(S^zk_x)=\lambda_W/\lambda_C$ for $k_y=0$. Therefore, we can fit the spin texture by averaging the $S^y/S^x$ on the $k_x$ axis and $S^xk_z/(S^zk_x)$ in the $k_z$-$k_x$ plane. Note that the second conduction and valence bands are hybridized with the deeper bands close to the Brillouin zone boundary, see Fig. 2(c). So, we choose a smaller range in the $k$ space to fit the spin texture, as labeled in the yellow dashed frame in Fig. 6 (a,d).

The model fitted spin textures are shown in Fig. S5 (e-h) for the $k_x$-$k_y$ plane and Fig. 6 (e-h) for the $k_z$-$k_x$ plane. For the second conduction band in Fig. 5,6(e), we get $\lambda_W=-\,0.6\lambda_C$, $\lambda_R=-\,1.6\lambda_C$, where $\lambda_C=C$ ($C$ is a constant). For the first conduction band in Fig. 5,6(f), we get $\lambda_W=-\,0.5\lambda_C$, $\lambda_R=-\,1.2\lambda_C$. For the first valence band in Fig. 5,6(g), we get $\lambda_W=1.5\lambda_C$, $\lambda_R=0.7\lambda_C$. For the second valence band in Fig. 5,6 (h), we get $\lambda_W=1.7\lambda_C$, $\lambda_R=1.1\lambda_C$. We find that the spin texture cannot fit well close to the Brillouin boundaries, which is due to the spin degeneracy at the boundaries. The highly degenerate band structure for the high symmetry lines at the Brillouin boundaries is shown in Fig. S7(a). The corresponding reciprocal lattice is displayed in Fig. S7 (b).

Next, we discuss the monolayer system. Fig. S8(a) shows the structure of the (100) surface of the chiral bulk in Fig. 2(a), and the corresponding band structure is shown in Fig. S7(c). The spin texture of the first valence band in Fig. 3(f) is reproduced in Fig. S8(b) for convenience. For monolayer, $S^y/S^x=(\lambda_{D'}-\,\lambda_R)/(\lambda_D+\lambda_W)$ for $k_y=0$. By fitting the spin texture around $\Gamma$ point, we get $\lambda_D=-\,2.7\lambda_C$ and $\lambda_{D'}=0.4\lambda_C$ assuming the SOC strength for $\lambda_W=1.5\lambda_C$ and $\lambda_R=0.7\lambda_C$ are unchanged. The fitted spin texture is shown in Fig. S8(c). We also investigate the monolayer of the (110) surface of the achiral bulk in Fig. 4(d), as shown in Fig. S8(d). By fitting the spin texture around $\Gamma$ point, we get $\lambda_D=-\,1.9\lambda_C$, $\lambda_{D'}=0.7\lambda_C$. Compared with the (100) monolayer, although $\lambda_D$ is decreased by 30\%, the $\lambda_{D'}$ is increased by 75\%. As shown in Fig. S8 (d), the inter-chain I atoms on the (110) monolayer form a tetrahedra shape, which may result in the increase of $\lambda_{D'}$. The enhanced $\lambda_{D'}$ leads to an opposite $S^y$ for the same $k$, comparing Fig. S8 (c) and (f).

Except for thickness reduction to monolayer, we also apply a compressive strain to reduce symmetry. As shown in Fig. S9, the compressive strain is applied along the $x$ direction with 0\%, 5\%, and 10\%, respectively. From the spin texture of the first valence band, we can observe that the spins along $k_x$ reverse directions first from the center of the Brillouin zone [see Fig. 9(b)], then gradually to the whole Brillouin zone [see Fig. 9(c)]. In experiments, there are more methods to reduce symmetry, for example, applying an external electric or magnetic field, introducing impurities or vacancies, and modifying the shape or composition of the material.

\clearpage
\begin{figure}[H]
	\centering
	\includegraphics[width=0.6\columnwidth]{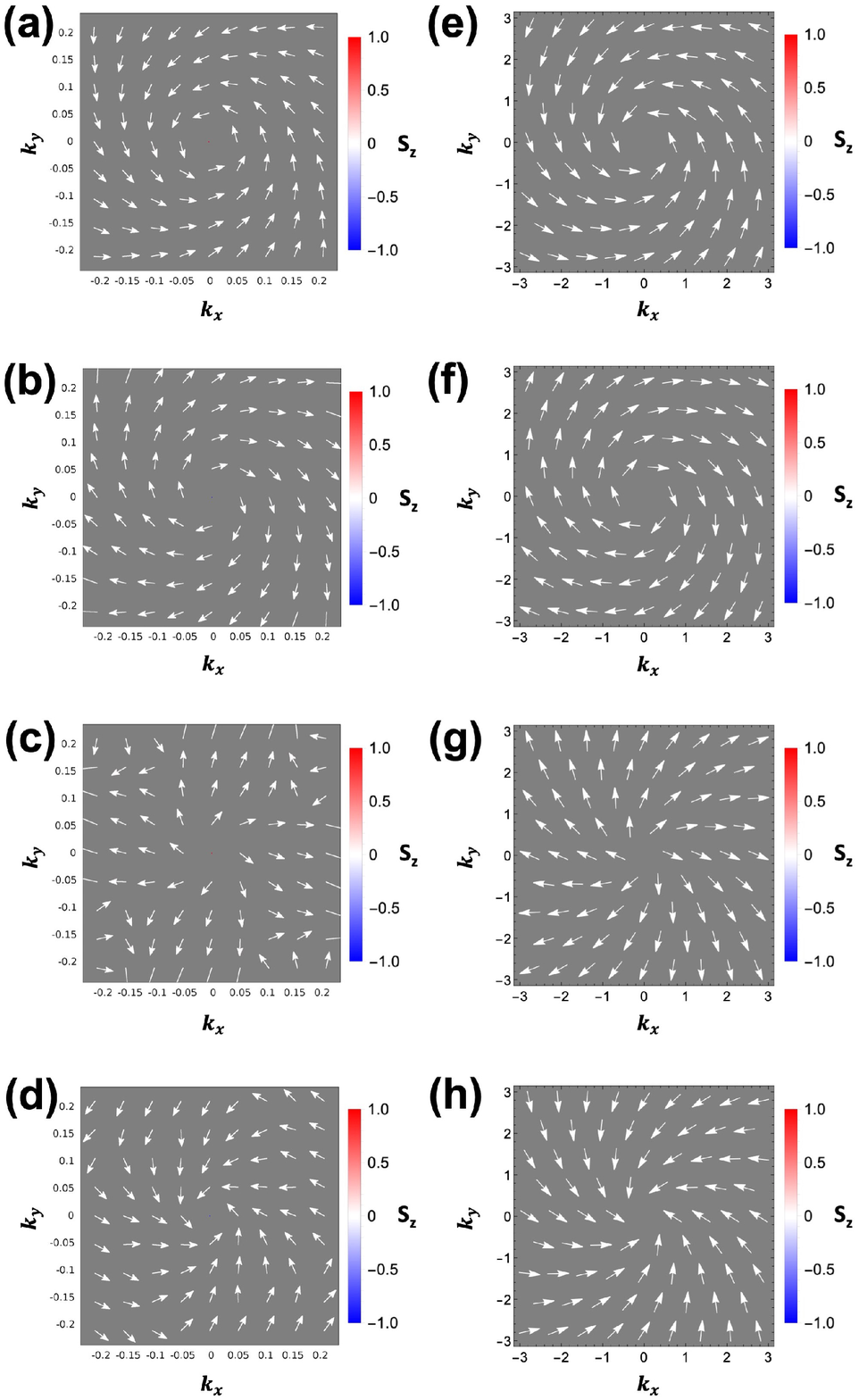}
	\caption{\label{fig:epsart} The spin texture of the bulk InSeI in the $k_x$-$k_y$ plane for (a) the second conduction band, (b) the first conduction band, (c) the first valence band (d) the second valence band. (e-h) The model fitted spin texture of (a-d).}
\end{figure}

\begin{figure}[H]
	\centering
	\includegraphics[width=0.6\columnwidth]{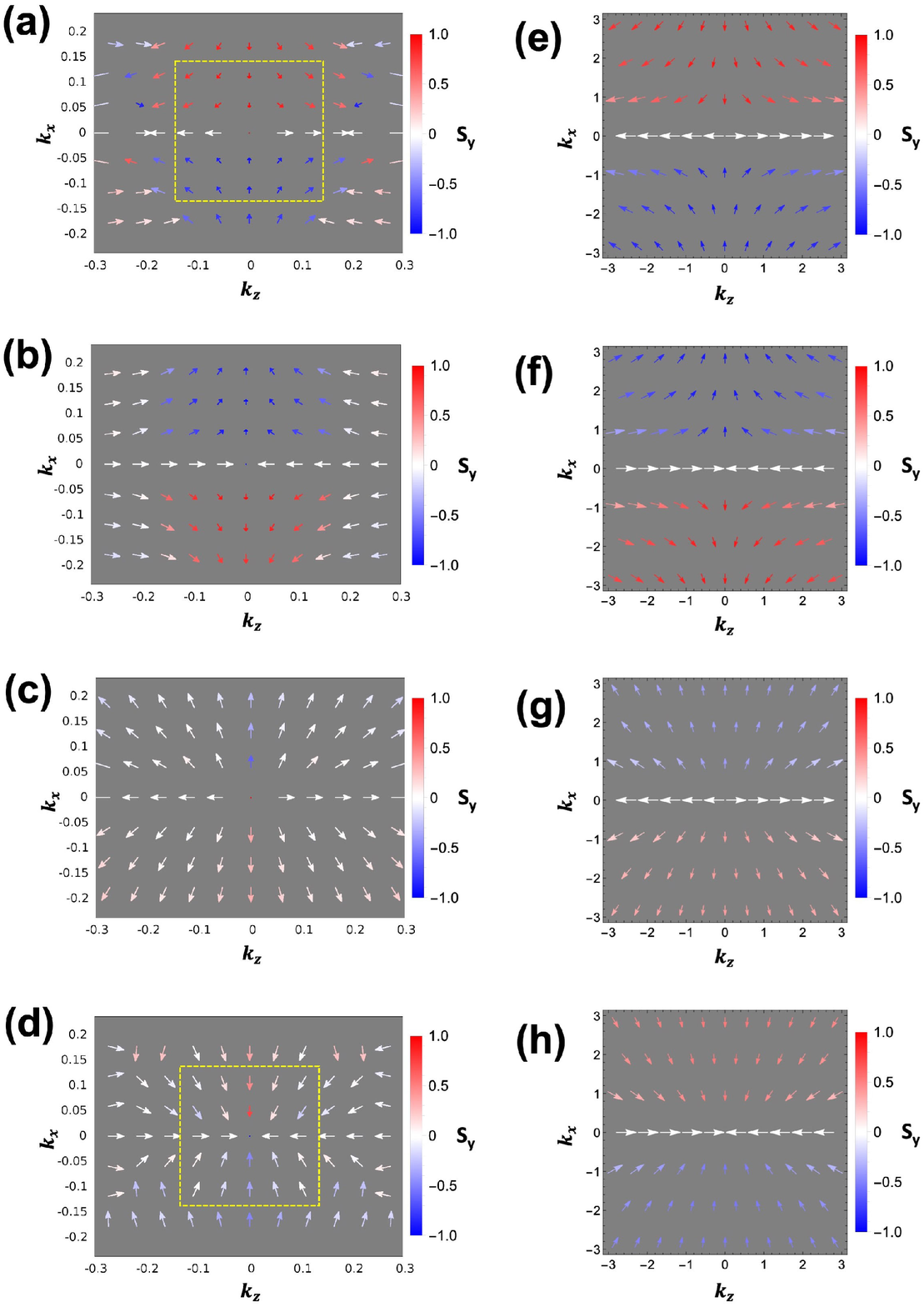}
	\caption{\label{fig:epsart} The spin texture of the bulk InSeI in the $k_z$-$k_x$ plane for (a) the second conduction band, (b) the first conduction band, (c) the first valence band (d) the second valence band. (e-h) The model fitted spin texture of (a-d). The yellow dashed frame represents the fitted data range for the second bands.}
\end{figure}

\begin{figure}[H]
	\centering
	\includegraphics[width=1\columnwidth]{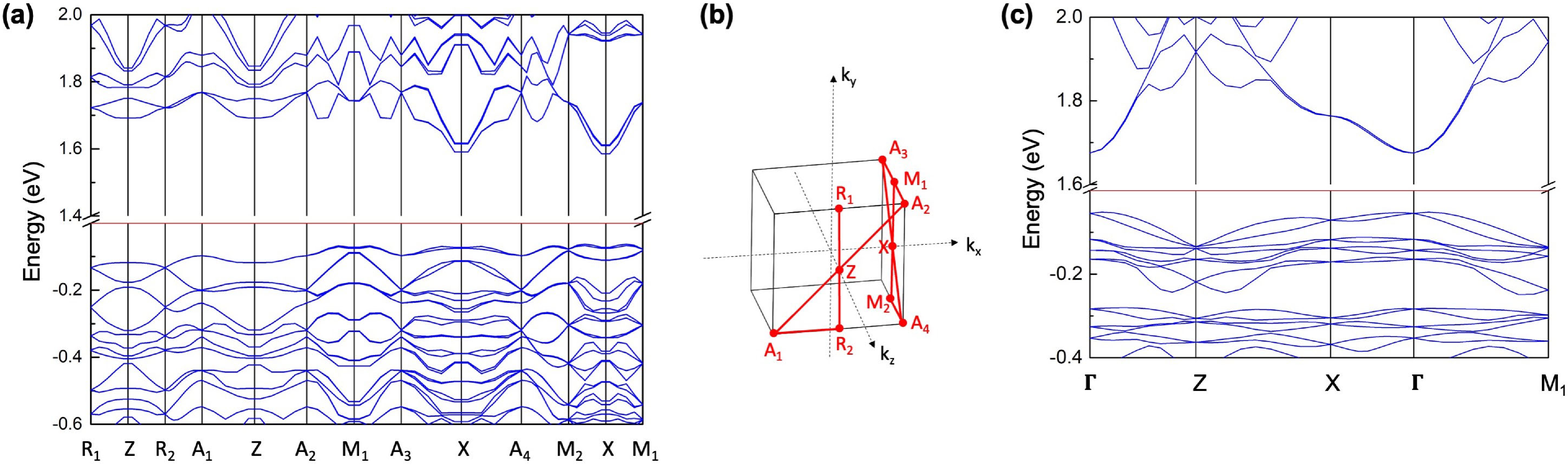}
	\caption{\label{fig:epsart} (a) The highly degenerate band structures for the high symmetry lines at the Brillouin boundaries for the chiral bulk InSeI. (b) The corresponding reciprocal lattice labeled with high symmetry points. The red lines correspond to the high symmetry lines in (a). (c) The band structures for the monolayer for the (100) surface of the chiral bulk.}
\end{figure}

\begin{figure}[H]
	\centering
	\includegraphics[width=0.6\columnwidth]{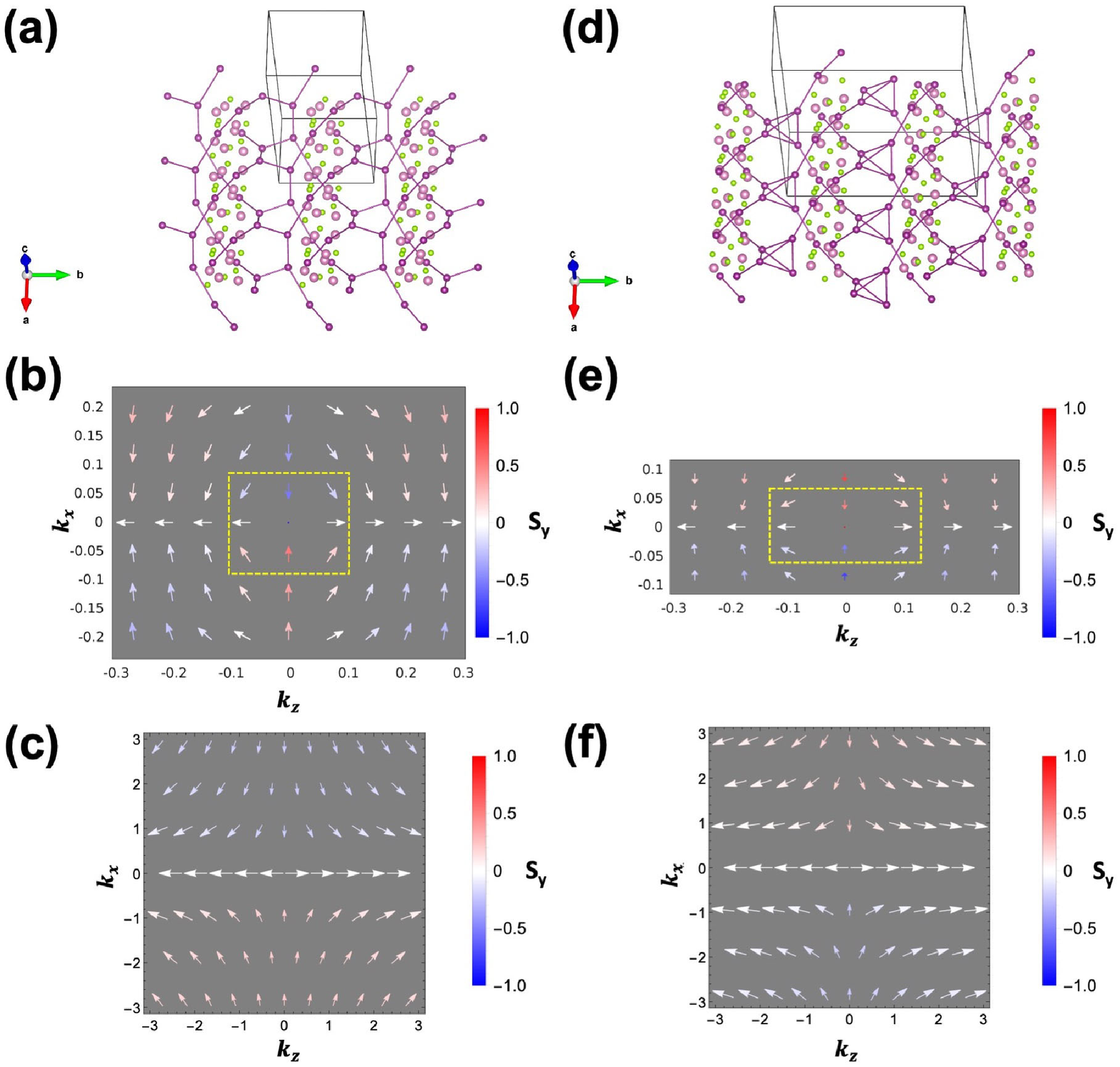}
	\caption{\label{fig:epsart} The lattice structure of (a) the (100) surface of the chiral bulk (c) the (110) surface of the achiral bulk. We connected the inter-chain I-I atoms to clarify the different nanochain alignments of the two monolayer structures. The spin texture (b,e) from the first-principle calculations and (c,f) from the model fitting for the (b,c) (100) and (e,f) (110) monolayer, respectively. The yellow dashed frame represents the fitted data range.}
\end{figure}

\begin{figure}[H]
	\centering
	\includegraphics[width=1\columnwidth]{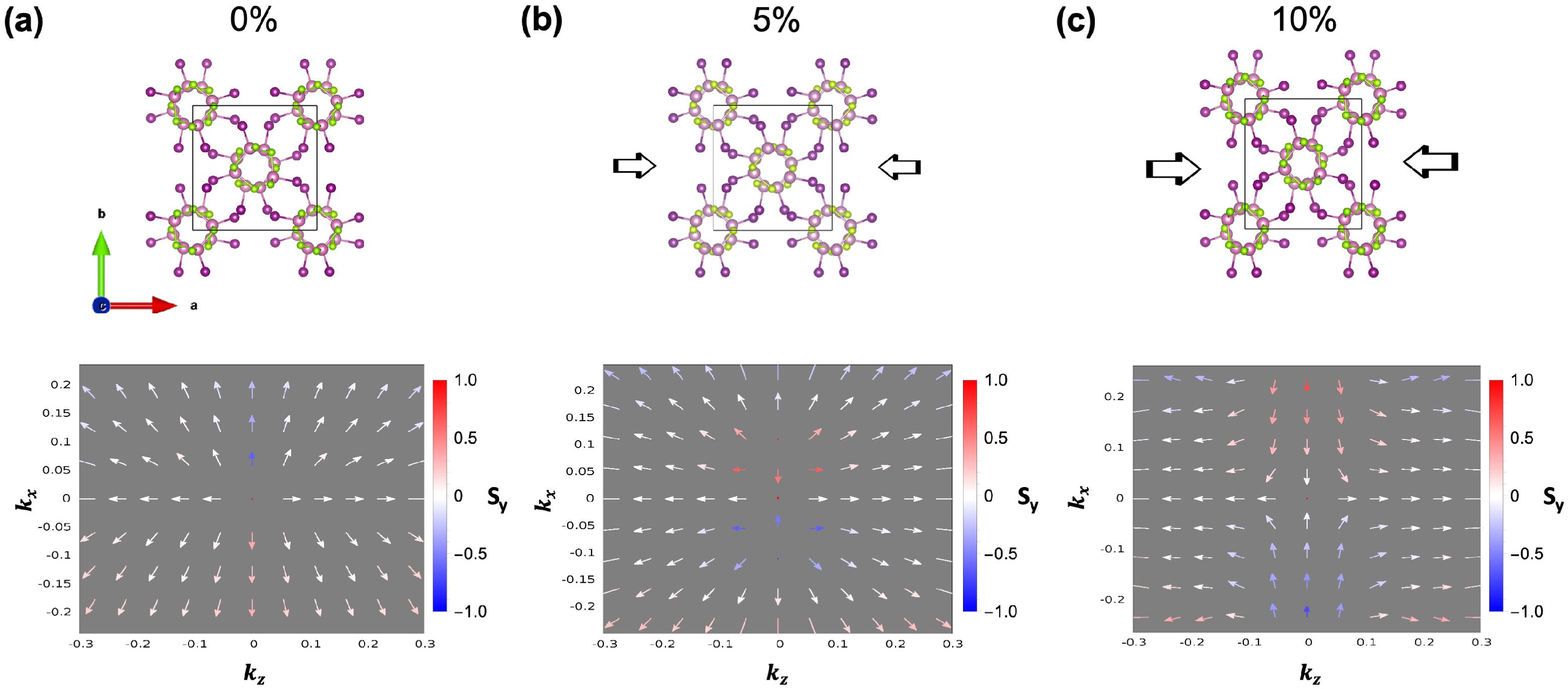}
	\caption{\label{fig:epsart} The lattice structure and spin texture of the first valence band with a compressive strain along $x$ direction. (a) the chiral bulk exhibits $C_4$ symmetry in the normal plane with 0\% strain. The symmetry is reduced to $C_2$ with (b) 5\% and (c) 10\% strains.}
\end{figure}

\section{Experimental method}

InSeI was grown by melt synthesis. In (STREM Chemicals, 99.9\%), Se (STREM Chemicals, 99.5\%), and I$_2$ (Spectrum Chemical, 99.8\%) in a 1:1:1 In:Se:I ratio were combined in a Pyrex ampoule and sealed under vacuum ($<$20 mTorr). The precursor mix was heated to 450$^\circ$C and held there for 4 days, then cooled down to room temperature for 1 day.

Scanning electron microscopy (SEM) images and energy-dispersive X-ray spectroscopy (EDS) analysis were performed on InSeI using FEI Quanta three-dimensional (3D) FEG at 20 keV working voltage. High-resolution transmission electron microscopy (HRTEM) images were collected on JEOL JEM-2800 TEM at 200keV working voltage and using the Metro In-situ Counting Camera, which was on loan from Gatan. The HRTEM sample was prepared by contacting the TEM grid onto PDMS gel with mechanically exfoliated InSeI crystals. Indexing of the Fast Fourier Transform (FFT) for the HRTEM image was done using SingleCrystal software. The structure of InSeI slab exfoliated at the (110) plane is plotted in Fig. S10.

\begin{figure}[H]
	\centering
	\includegraphics[width=0.35\columnwidth]{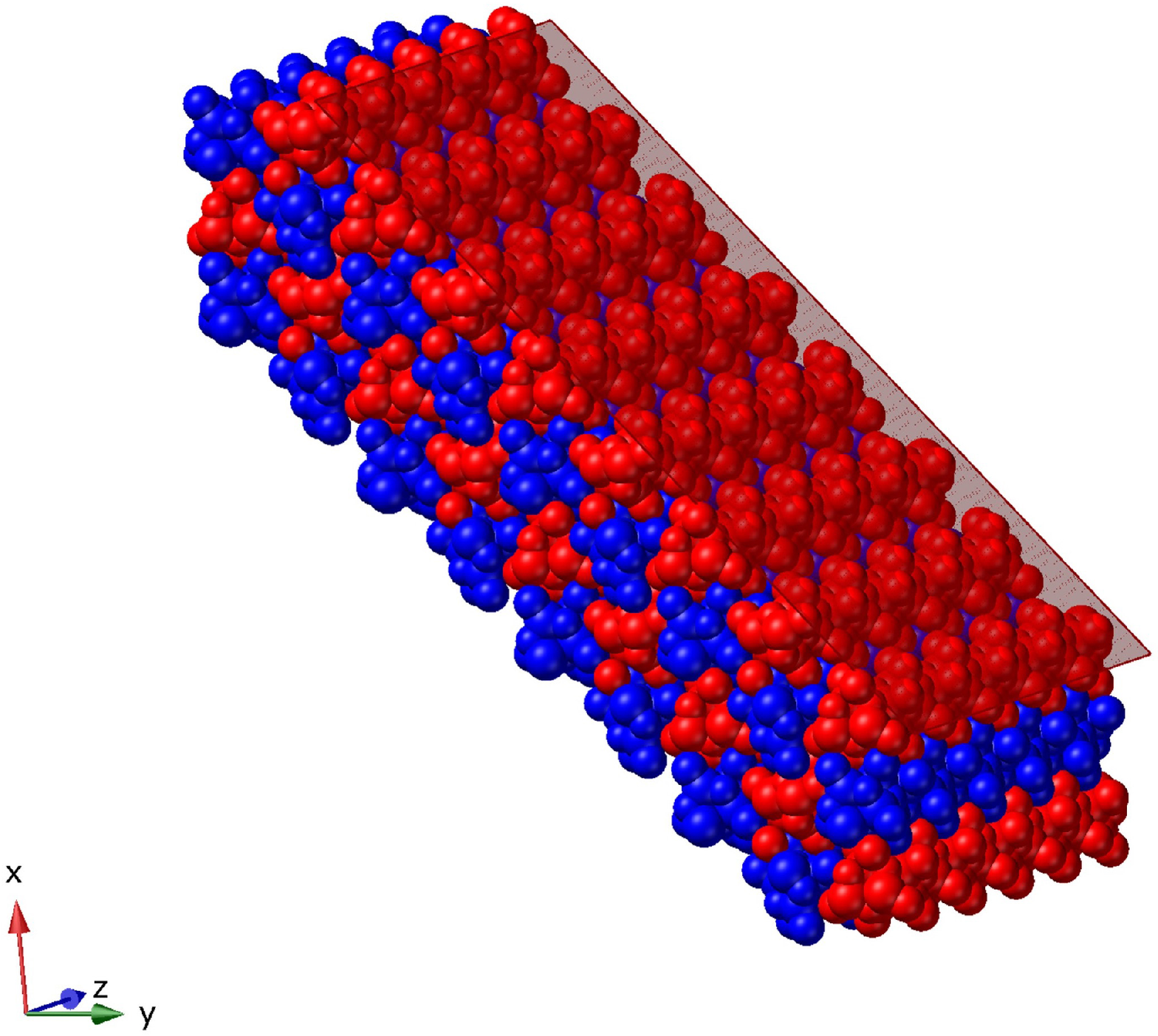}
	\caption{\label{fig:epsart} Schematic InSeI slab exfoliated at the (110) plane (highlighted in grey). Left- and right-handedness of nanochains are colorer in red and blue, respectively.}
\end{figure}

\section{Device application}

Last but not least, we propose a spintronic device architecture that can potentially probe spin selectivity in three directions, as shown in Fig. S11. Recent studies employed a similar device configuration with only one source and voltage meter ($S_1$ and $V_1$ in Fig. S11) to detect CISS of a chiral crystal \cite{inui2020chirality}. Based on the inverse spin Hall effect \cite{valenzuela2006direct,saitoh2006conversion,kimura2007room}, the spin polarization can be converted to the charge current by a tungsten electrode with a large spin Hall angle \cite{wang2014scaling,niimi2015reciprocal}. As we have previously described and discussed, our proposed hypothetical InSeI structure displays spin selectivity across three dimensions. As a result, it is possible to detect spin polarization not only along the chiral direction ($S_1$ and $V_1$) but also along the perpendicular directions ($S_2$, $S_3$ and $V_2$, $V_3$ in Fig. S11). In other words, more complex three-dimensional operations for a spintronic device can be realized. We postulate that this multi-axial spintronic device architecture can store and process information at a higher density and may also represent a quaternary or even octal numeral system.

\begin{figure}[H]
	\centering
	\includegraphics[width=0.5\columnwidth]{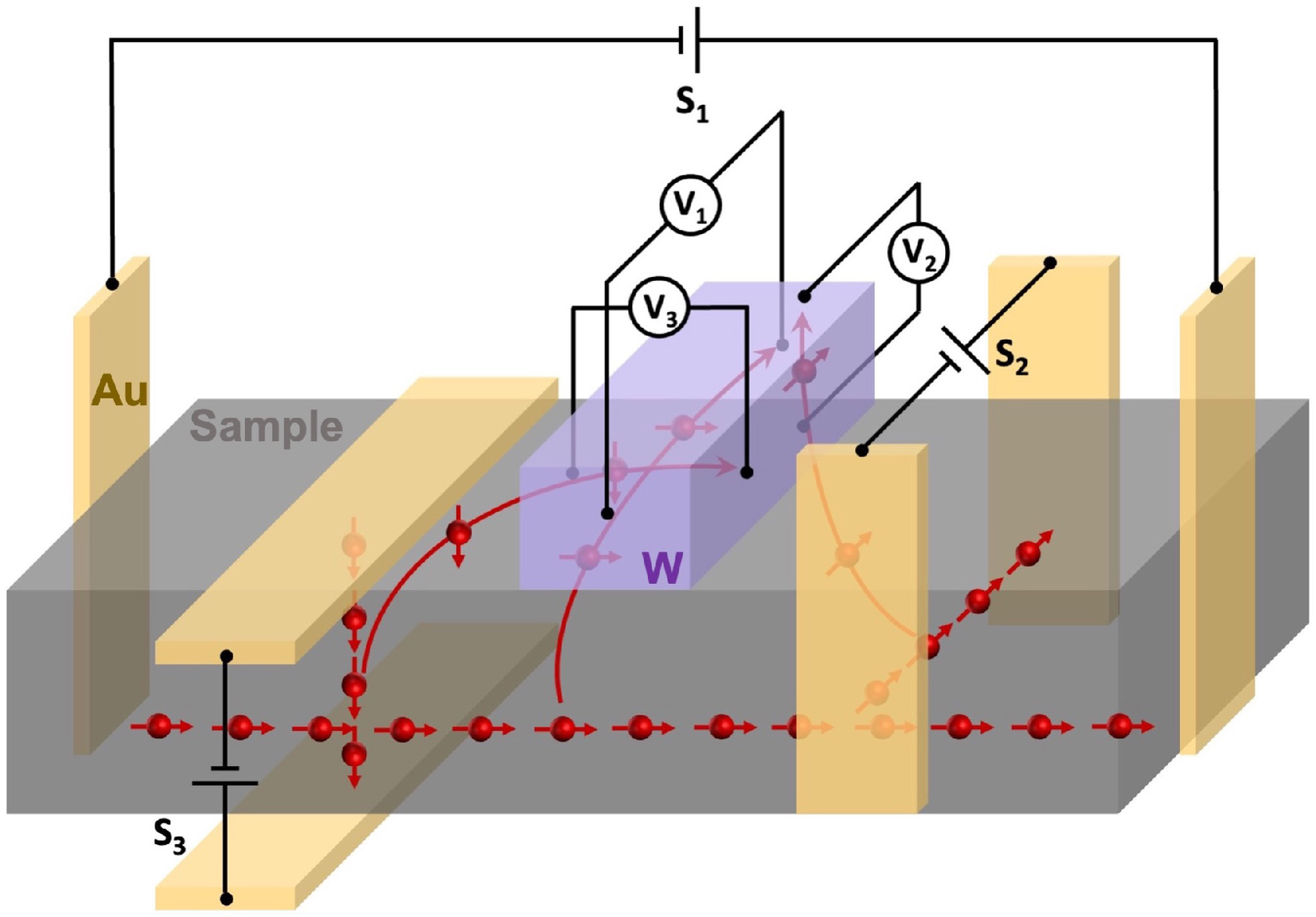}
	\caption{\label{fig:epsart} The schematic figure of the spintronic device to measure spin selectivity in three directions. $S_{1,2,3}$ are three independent sources and $V_{1,2,3}$ are three voltage meters to measure the charge current in three directions. The grey cube represents InSeI sample, and the purple and gold cubes represent tungsten (W) and gold (Au) electrodes.}
\end{figure}

\clearpage

\end{document}